\documentclass[aps,preprint,endfloats*,amsfonts,amsmath]{revtex4}

\usepackage{graphicx}  

\begin{document}
\title{Bilayers in amphiphilic mixtures connected by threadlike micelles: a self-consistent field theory study}
\author{Martin J.~Greenall}
\affiliation{Theoretical Soft Matter and Biophysics, Institute for Complex Systems,
  Forschungszentrum J\"{u}lich, 52425 J\"{u}lich, Germany}
\email{mjgreenall@physics.org}
\author{Gerhard Gompper}
\affiliation{Theoretical Soft Matter and Biophysics, Institute for Complex Systems,
 Forschungszentrum J\"{u}lich, 52425 J\"{u}lich, Germany}
\begin{abstract}
Binary mixtures of amphiphiles in solution can self-assemble into a wide range
of structures when the two species
individually form aggregates of different curvatures. A specific example of
this is seen in solutions of lipid mixtures where the two species
form lamellar structures and spherical micelles respectively. Here,
vesicles connected by thread-like micelles can form in a narrow
concentration range of the sphere-forming lipid. We present a
self-consistent field theory (SCFT) study of these structures. Firstly, we
show that the addition of sphere-forming lipid to a solution of
lamella-former can lower the free energy of cylindrical, thread-like
micelles and hence encourage their formation. Next, we demonstrate the
coupling between composition and curvature; specifically,
that increasing the concentration of sphere-former in a system of two
bilayers connected by a thread leads to a transfer of amphiphile to
the thread. We further show that the two species are segregated within
the structure, with the concentration of sphere-former being
significantly higher in the thread. Finally, the addition of larger amounts of sphere-former
is found to destabilize the junctions linking the bilayers to the
cylindrical micelle, leading to a breakdown of the connected
structures. The degree of segregation of the amphiphiles and the
amount of sphere-former required to destabilize the junctions is shown
to be sensitive to the length of the hydrophilic block of the sphere-forming amphiphiles.
\end{abstract}
\maketitle
\section{Introduction}

When dissolved in solution, amphiphiles such as lipids and block
copolymers can self-assemble into a remarkably wide range
of different structures
\cite{jain_bates,battaglia_ryan}. This phenomenon
has been the subject of much recent research \cite{smart,howse},
especially given the potential applications of self-assembled
amphiphile aggregates in the encapsulation and delivery of drugs and
genetic material \cite{kim,lomas} . 

In the case of dilute solutions of a single type of simple amphiphile, it is often relatively
straightforward to understand, at least on a qualitative level, why a given structure
forms in a given system \cite{kinning_winey_thomas}. The major
factor in determining the shape of the aggregates is the architecture
of the amphiphile; that is, the size of hydrophilic and hydrophobic components. If the hydrophilic component is large compared to the hydrophobic component, then curved aggregates such as spherical or cylindrical micelles form. The hydrophilic component can be intrinsically large (like a bulky head-group) or can appear large due to interaction with the solvent (as for a polymer chain in good solvent). Conversely, if the hydrophobic component is large, lamellar structures such as vesicles are observed.

The picture becomes far more complex when two types of amphiphile
that individually form different structures are
mixed \cite{jain_bates}. In such systems, new structures not seen in
monomodal dispersions are observed, such as 
undulating cylinders \cite{jain_bates}. The current interest in such
amphiphile blends \cite{schuetz,sorre} has two major
motivations. The first of these is simply that the presence of two amphiphiles increases the number
of design parameters available for the control of the self-assembly
process. For example, the architectures of both amphiphiles may now be varied, as may
the amount of each species. Less obviously, the blending process
itself may be manipulated; in particular, the amphiphiles may be mixed
together before, during, or after their individual self-assembly
\cite{schuetz}. The second factor driving research into blends of
amphiphiles is their importance in biological systems. Cells contain
mixtures of lipids, which are transported between the various
compartments of the cell by mechanisms not yet fully understood \cite{akiyoshi}. A
topic of current debate is the role of membrane curvature in the
transport process, and in particular in the sorting of different lipid
species within the cell \cite{sorre}. A model system ideally suited to
isolate and investigate the links between curvature and lipid segregation is a
large vesicle (of low curvature) attached to a long, thread-like tube
(of high curvature) \cite{sorre,heinrich}. Examination of the
distribution of the different lipid components within this structure
then yields information on the connections between membrane curvature,
composition, and sorting.

Although this structure is often investigated through the manipulation of
individual vesicles using micropipettes \cite{heinrich}, recent
experiments performed by Zidovska {\em et al.} have found similar
aggregates consisting of two vesicles connected by thread-like
micelles or hollow bilayer-membrane cylinders \cite{zidovska} in bulk solutions of
lipid mixtures in a narrow concentration regime. In these experiments, the curvature-stabilizing amphiphiles have an extremely large, branched headgroup, which may account for the fact that threads and tubes form spontaneously here, rather than having to be pulled mechanically from a vesicle (although some mechanical stress produced by flow or by the vesicle formation process may still be present). These authors also explain their results in terms of lipid sorting, and propose a
pictorial model of connected vesicles in which the thread contains
higher concentrations of curvature-stabilizing amphiphiles than the
two flatter bilayers it connects. 

We investigate here how much information about connected bilayers can be
obtained from a simple mean-field model of two types of amphiphile in
solution, without considering details of the shape of the
individual lipids or more complex effects such as lipid clustering \cite{sorre}. In
particular, we will investigate the coupling between
composition and shape; that is, how addition of a
micelle-forming species changes the balance between regions of high
and low
curvature in the aggregate. The segregation of the different lipid species
between these regions will also be studied. Furthermore, we will consider the stability
of the junction between the bilayer and a thread, and how this is
affected by the addition of micelle-forming amphiphile. Finally, we
will consider the effect of varying the architecture of the
micelle-forming additive on the above phenomena. To study these problems in as simple a form as possible, we will consider two amphiphiles that have no interactions leading to phase separation of the different species.  To this end, the amphiphiles will be formed of the same two materials A and B but will individually favor different curvatures as a result of the lengths of their hydrophobic and hydrophilic blocks.

The paper is organized as follows. In the following section, we
introduce the theoretical technique to be used (self-consistent field
theory) and discuss its suitability for the current problem and what
information it can yield. We then present and discuss our theoretical results,
and give our conclusions in the final section.

\section{Self-consistent field theory}\label{scft}

Self-consistent field theory (SCFT) \cite{edwards} has been used
successfully over a number of years to model the equilibrium structures formed in
melts and blends of polymers \cite{maniadis,drolet_fredrickson,matsen_book}, and may also be used to investigate
metastable structures \cite{duque,katsov1}. Several features of SCFT
recommend it as a method to address the current problem of connected
bilayers. Firstly, its general
advantages are that it requires significantly less computational power than simulation
methods such as Monte Carlo, yet often provides comparably accurate
predictions of the form of individual aggregates
\cite{cavallo,wijmans_linse,leermakers_scheutjens-shape}. Secondly, as a
relatively simple model, with a coarse-grained description of the polymer molecules, it will allow us to capture the
basic phenomenology of the system clearly and give insight into
how general the phenomena observed are likely to be. In
addition, the theory possesses two specific features that are
especially advantageous in the current problem. The first of these is that the numerical
algorithm that we use to solve the SCFT equations requires an initial guess for
the form of the solution. This may be chosen so that solutions of a
particular shape, even a rather complex one such as the connected bilayer structure of interest
here, may be sought. This is a more direct and straightforward method of
systematically studying a given structure than waiting for it to form
by chance in a dynamical simulation. The second specific advantage of
SCFT is that it makes no initial assumption about the segregation of two
structurally different amphiphiles within the self-assembled
aggregate, provided the two amphiphiles are formed from the same types
of monomer. This will enable us to demonstrate that any such
segregation effects
arise spontaneously and need not be `added by hand' to the theory.

To illustrate some of the points made above, and for the sake of
concreteness, we now outline briefly the assumptions and mathematical structure of
SCFT, and discuss its implementation in the current system. The theory models individual amphiphiles as random walks
in space: fine details of their structure and packing are neglected \cite{schmid_scf_rev}. An ensemble
of many such molecules is considered. The interactions between the molecules are
modeled on a comparably simple level: by assuming that the blend is incompressible and introducing a
contact potential between the molecules \cite{matsen_book}. The strength of this
potential is specified by the Flory parameter $\chi$ \cite{jones_book}.

The first step in finding an approximation method (SCFT) for the
above system is to view each molecule as being acted on by a field
produced by all other molecules in the system \cite{matsen_book}.
This way of looking at the problem, which as yet involves no approximations, has several advantages when
computing numerical solutions. Firstly, it transforms the $N$-body problem of modeling an ensemble of $N$
polymers into $N$ $1$-body problems \cite{matsen_book}. Since we consider the
partition sum over all possible system configurations, all molecules
of a given species may be treated as equivalent and we need
only solve one $1$-body problem for each molecular species in the
system. Secondly, this approach allows us to convert the discrete sum
over polymer configurations into an integral over smooth functions,
which is easier to treat numerically \cite{katsov1}. Finally, the computational
difficulty of the problem may be sharply reduced by finding approximate forms of
the field variables using a saddle-point approximation \cite{schmid_scf_rev}. These
approximations correspond to neglecting fluctuations in the system.

SCFT can be used to study not only simple homopolymers, but also diblock and more complex copolymers \cite{mueller} and any given mixture of these \cite{denesyuk}. We now discuss the application of SCFT to our system of two
amphiphiles in a solvent, which we model by a simple
mixture of two types of block copolymer with a homopolymer solvent. Although
considerably simpler than the experimental lipid system we wish to
consider \cite{zidovska}, such models
have been successfully used to study biological amphiphilic systems
\cite{katsov1}, and can isolate individual phenomena more clearly than
more complex theories. We take the lamella-forming species of
copolymer to have a mean-squared
end-to-end distance of $a^2N$, where $a$ is the monomer length and $N$
is the degree of polymerization \cite{matsen_book}. One half of the
monomers in this polymer are hydrophilic (type A) and the other half are hydrophobic
(type B). For simplicity, we choose the same value of $a^2N$ for the A
homopolymer solvent. The curvature-stabilizing species has the same
size hydrophobic block as the lamella-former, but necessarily has a
longer hydrophilic block. We consider three
different mean-square end-to-end distances $\alpha 
a^2N$ for this species, with $\alpha=1.5$, $2$ and $3$.

In this paper, we keep the amounts of copolymer and homopolymer in the
simulation box fixed; that is, we work in the canonical ensemble. This
will make it easier for us to access more complex metastable
structures, such as the connected bilayer, by a
suitable choice of ansatz for the SCFT equations. Such structures are
more difficult to stabilize in ensembles where
the system is able to relax by
varying the amount of the various species, and can require the
imposition of detailed constraints on the final shape of the aggregate
\cite{katsov1}, a step we wish to avoid in our current study of the
relationship between composition and curvature.

Applying the procedure described above, we find that the SCFT approximation to the free energy of our system has the form
\begin{eqnarray}
\lefteqn{\frac{FN}{k_\text{B}T\rho_0V}=\frac{F_\text{h}N}{k_\text{B}T\rho_0V}}\nonumber\\
& &
-\chi N(\phi_\text{A}(\mathbf{r})+\phi_\text{A2}(\mathbf{r}) +\phi_\text{S}(\mathbf{r})-\overline{\phi}_\text{A}-\overline{\phi}_\text{A2}-\overline{\phi}_\text{S})(\phi_\text{B}(\mathbf{r})+\phi_\text{B2}(\mathbf{r})-\overline{\phi}_\text{B}-\overline{\phi}_\text{B2})
\nonumber\\
& &
-(\overline{\phi}_\text{A}+\overline{\phi}_\text{B})\ln (Q_\text{AB}/V)-[(\overline{\phi}_\text{A2}+\overline{\phi}_\text{B2})/\alpha]\ln
(Q_\text{AB2}/V) -\overline{\phi}_\text{S}\ln (Q_\text{S}/V)
\label{FE}
\end{eqnarray}
where the $\overline{\phi}_i$ are the mean volume fractions of the
various components. The $\phi_i(\mathbf{r})$
are the local volume fractions, with $i=A$ or $A2$ for the hydrophilic components of species $1$ and $2$, $i=B$ or $B2$ for the hydrophobic components and $i=S$ for the solvent. The strength of the repulsion between the species A (hydrophilic component and solvent) and B (hydrophobic component) is determined by the Flory parameter $\chi$. $V$ is the
total volume, $1/\rho_0$ is the volume of a monomer, and $F_\text{h}$
is the SCFT free energy of a homogeneous system containing the same
components. The details of the individual polymers enter through the
single-chain partition functions $Q_i$. As an example, that for the
homopolymer is given by \cite{matsen_book}
\begin{equation}
Q_\text{S}[W_\text{A}]=\int\mathrm{d}\mathbf{r}\,q_\text{S}(\mathbf{r},s)q_\text{S}^\dagger(\mathbf{r},s)
\label{single_chain_partition}
\end{equation}
where the $q$ and $q^\dagger$ terms are single chain propagators
\cite{matsen_book}. The partition functions of the copolymer chains
are determined similarly. We now recall that the
polymer molecules are modeled as random walks subject to an external
field that incorporates their interactions with the rest of the
melt. This is reflected in the fact that the propagators satisfy
modified diffusion equations. Again considering the case of the homopolymer, we
have
\begin{equation}
\frac{\partial}{\partial
s}q_\text{S}(\mathbf{r},s)=\left[\frac{1}{6}a^2N\nabla^2-W_\text{A}(\mathbf{r})\right]q_\text{S}(\mathbf{r},s)
\label{diffusion}
\end{equation}
where $s$ is a curve parameter describing the position along the
polymer backbone and the initial condition is $q_\text{S}(\mathbf{r},0)=1$. The copolymer
propagators are computed in a similar way, with the copolymer
architecture entering into the theory through the fact that the
corresponding diffusion equation for the copolymer is solved with
the field $W_i(\mathbf{r})$ and the prefactor of the $\nabla^2q$
term appropriate to each of the two sections of the copolymer
\cite{fredrickson_book}. In the case of the longer sphere-forming
copolymer, the fields must be multiplied by the ratio $\alpha$ of the two
degrees of polymerization \cite{matsen_book}.

The derivation of the SCFT free energy $F$ also generates a set of simultaneous equations linking
the values of the fields and densities at the minimum. The first of
these is a simple statement of the incompressibility of the system;
however, we also find the following linear relation
\begin{equation}
w_\text{A}(\mathbf{r})-w_\text{B}(\mathbf{r})=2\chi N[\overline{\phi}_\text{A}+\overline{\phi}_\text{A2}+\overline{\phi}_\text{S}-\phi_\text{A}(\mathbf{r}) -\phi_\text{A2}(\mathbf{r})-\phi_\text{S}(\mathbf{r})]
\label{SCFT_equation}
\end{equation}
Furthermore, the homopolymer density is related to the propagators (see Equation \ref{diffusion}) according to
\cite{matsen_book}
\begin{equation}
\phi_\text{S}(\mathbf{r})=\frac{V\overline{\phi}_\text{S}}{Q_\text{S}[w_\text{A}]}\int^1_0\mathrm{d}s\,
q_\text{S}(\mathbf{r},s)q_\text{S}^\dagger(\mathbf{r},s)
\label{density}
\end{equation}
The copolymer densities are calculated similarly, with the
integration limits set to give the correct proportions of each
species.

In order to calculate the SCFT density profiles for a given set of
mean volume fractions, equation
\ref{SCFT_equation} must be solved with the densities calculated as
in Equation \ref{density} and respecting the incompressibility of the system. To begin, we make a guess
for the form of the fields $w_i(\mathbf{r})$ with the approximate form
of the structure we wish to study and solve the diffusion
equations to calculate the propagators and hence the densities
corresponding to these fields (see Equations \ref{diffusion} and
\ref{density}). New values for the fields are now calculated using
the new $\phi_i(\mathbf{r})$, and the $w_i$ are updated
accordingly \cite{matsen2004}. The procedure is repeated until convergence is achieved.

\begin{figure}
\includegraphics[width=0.5\linewidth]{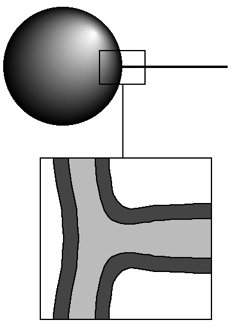}
\caption{\label{junction_fig} Schematic of a vesicle connected to a cylindrical micelle. In our calculations, we focus on the junction region sketched in the inset. Here, a vertical cut through the bilayer-cylinder junction is shown. The light region is the hydrophobic core of this structure, and the dark gray region is the hydrophilic corona. In Figures \ref{A_fig} and \ref{A4_fig}, we plot the density profile of the hydrophilic corona in order to demonstrate how the structure changes as sphere-forming amphiphile is added to the system. }
\end{figure}

The diffusion equations are solved using a finite difference method
\cite{num_rec} with step size of $0.04\,aN^{1/2}$. To study the
connected vesicle structure, we focus on a region near the junction of the vesicle and the thread-like micelle (Figure \ref{junction_fig}). The curvature of the vesicle is taken to be sufficiently small that it can be modeled by a flat bilayer. We also assume cylindrical symmetry about the
axis of the thread-like micelles, and hence
consider an effectively two-dimensional problem in a cylindrical
calculation box. Reflecting boundary
conditions are imposed at the edges of the system, so that we are in effect studying a system of two bilayers facing each other and connected by a thread-like micelle. SCFT is then used to calculate the density profiles of the different species in the system (see inset to Figure \ref{junction_fig}).

In addition,
we perform effectively one-dimensional calculations on spherical micelles,
infinite cylinders and infinite bilayers, to complement our
calculations on the connected bilayer by demonstrating how the free
energy of simple structures with different curvature changes as the
volume fraction of micelle-former is varied. At each copolymer
composition, the density profile and free energy of the optimum
(lowest free energy)
sphere, cylinder and bilayer are calculated. The free energies are
then plotted to produce a simple phase diagram as a function of the
volume fraction of micelle former.

The calculation of the optimum aggregates proceeds as follows \cite{gbm_macro,gbm_jcp}. Firstly, we
calculate the free-energy
density of a single spherical, cylindrical or planar aggregate at fixed volume
fraction using SCFT. The volume of the simulation box
containing the aggregate is then varied until the free-energy density is
minimized. This provides a simple model of a larger system (of fixed
volume and fixed copolymer volume fraction) containing many
aggregates, since such a system can minimize its
free energy by varying the number of aggregates and hence the volume
occupied by each. Although this approach can be used to
perform detailed calculations on micelle formation and shape
transitions \cite{gbm_macro,gbm_jcp}, we use it here to give a demonstration of how copolymer architecture and blending affect the
curvature of the self-assembled structures.

\section{Results and discussion}\label{results}
To begin, we calculate the free-energy densities of ideal spheres,
infinite cylinders and infinite bilayers using the method of variable subsystem size described above, and determine how these vary as the volume fraction of
curvature-stabilizing amphiphile is increased. We fix the overall volume fraction of
homopolymer to $10\,\%$. The Flory parameter is set to the moderate value of $\chi N=30$: this is large enough for the copolymers to self-assemble but not so large that all interfaces become sharp.  All free energies are
plotted with respect to that of the homogeneous solution with the same
composition; that is, we plot the quantity
$f=FN/k_\text{B}T\rho_0V-F_\text{h}N/k_\text{B}T\rho_0V$
(see equation \ref{FE}).

\begin{figure}
\includegraphics[width=\linewidth]{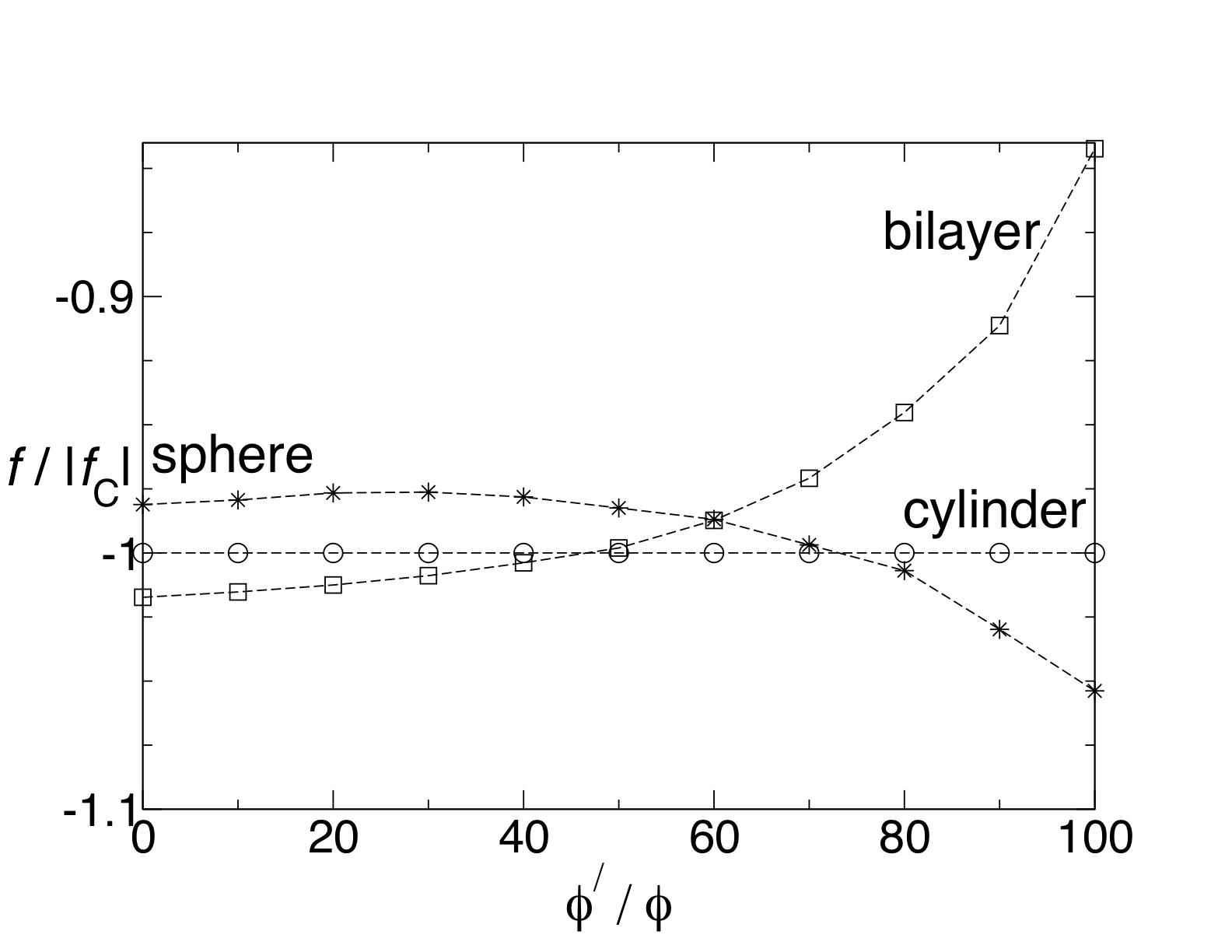}
\caption{\label{ms_fig} Free-energy densities of the
 optimum sphere (stars), cylinder (circles) and bilayer (squares) as
 the volume fraction $\phi'$ of sphere-former is increased at fixed
 overall volume fraction $\phi=10\,\%$.  In this case, the
 micelle-forming species is a strong sphere-former. Its mean-square
 end-to-end distance is $R_0^2=3a^2N$, and $1/6$ of its monomers are hydrophobic. The Flory parameter takes the moderate value of $\chi N=30$. As $\phi'$ is increased, the
 structure with the lowest free energy changes from the bilayer, then
 to the cylinder and finally to the sphere.}
\end{figure}

As discussed in the
preceding section, we use a symmetric lamella-former in
all cases, but blend this with different curvature-stabilizing
amphipiles. We first consider the
strongly sphere-forming amphiphile with mean-square end-to-end
distance $R_0^2=3a^2N$ and fraction of hydrophilic blocks $5/6$. 
Figure \ref{ms_fig} shows the free-energy densities of spherical, cylindrical
and planar aggregates as a function of $\phi'/\phi$, where $\phi'$ is
the volume fraction of sphere-forming amphiphile and $\phi$ is the
total volume fraction of amphiphile (sphere formers plus lamella
formers). The free-energy densities $f_i$ are quite close in
magnitude, and so are plotted normalized with respect to the
magnitude of the free-energy density of the cylinder $|f_\text{C}|$ in
order to show the shape transitions clearly. The cylinder
free-energy density thus appears as a horizontal line at $f=-1$, and
is approached from above and below by the sphere and lamella free
energy densities as the volume fraction of sphere former is
increased. At low sphere-former volume fractions, the lamellar
structure has the lowest free energy. At around $\phi'/\phi=50\,\%$,
the lamellar and cylindrical free energies cross, and the cylinder has the lowest free
energy until $\phi'/\phi\approx 70\,\%$, when the sphere finally
becomes most energetically favorable. There are two important points
to note from this calculation in the context of the current
problem. Firstly, we show that a blend of sphere-forming and
lamella-forming amphiphiles can preferentially form cylindrical structures (such as
the threads between vesicles in the experiments of Zidovska {\em et
 al}) \cite{zidovska}, even though neither of the amphiphiles forms this structure
individually. Secondly, at higher sphere former concentrations, the 
structures with positive curvature (spheres and cylinders) are shown
to have, by some distance, the lowest free energies. This point will be of relevance when
discussing the stability of the junctions between bilayers and threads
later. We note that the apparent small rise in the sphere-former free energy $f_\text{S}$
at low $\phi'/\phi$ is an artifact of normalizing the results with
respect to $|f_\text{C}|$. In the original units, the difference between
$f_\text{S}$ and $f_\text{C}$ is almost constant for small
$\phi'/\phi$. However, $|f_\text{S}|$ and $|f_\text{C}|$
decrease steadily as the amount of hydrophobic material in the system
decreases. Recalling that $|f_\text{S}|<|f_\text{C}|$, we see that the
result of this is that the ratio $|f_\text{S}|/|f_\text{C}|$ decreases slightly, and so
$f_\text{S}/|f_\text{C}|$ shows the small rise seen in Figure \ref{ms_fig}.

The shape transitions that occur as increasing concentrations of micelle-former are added to a solution of lamella-forming amphiphile have been studied in detail in the context of lipid-detergent mixtures \cite{vinson,oberdisse}, and the sequence of morphologies calculated above (vesicles, cylindrical micelles, spherical micelles) is indeed observed. These transitions have also been computed using lattice SCFT by Li {\em et al.} for a single set of amphiphile parameters \cite{li}. The bilayer-cylinder transition can also be obtained from a much simpler model of membrane curvature \cite{andelman}.

\begin{figure}
\includegraphics[width=\linewidth]{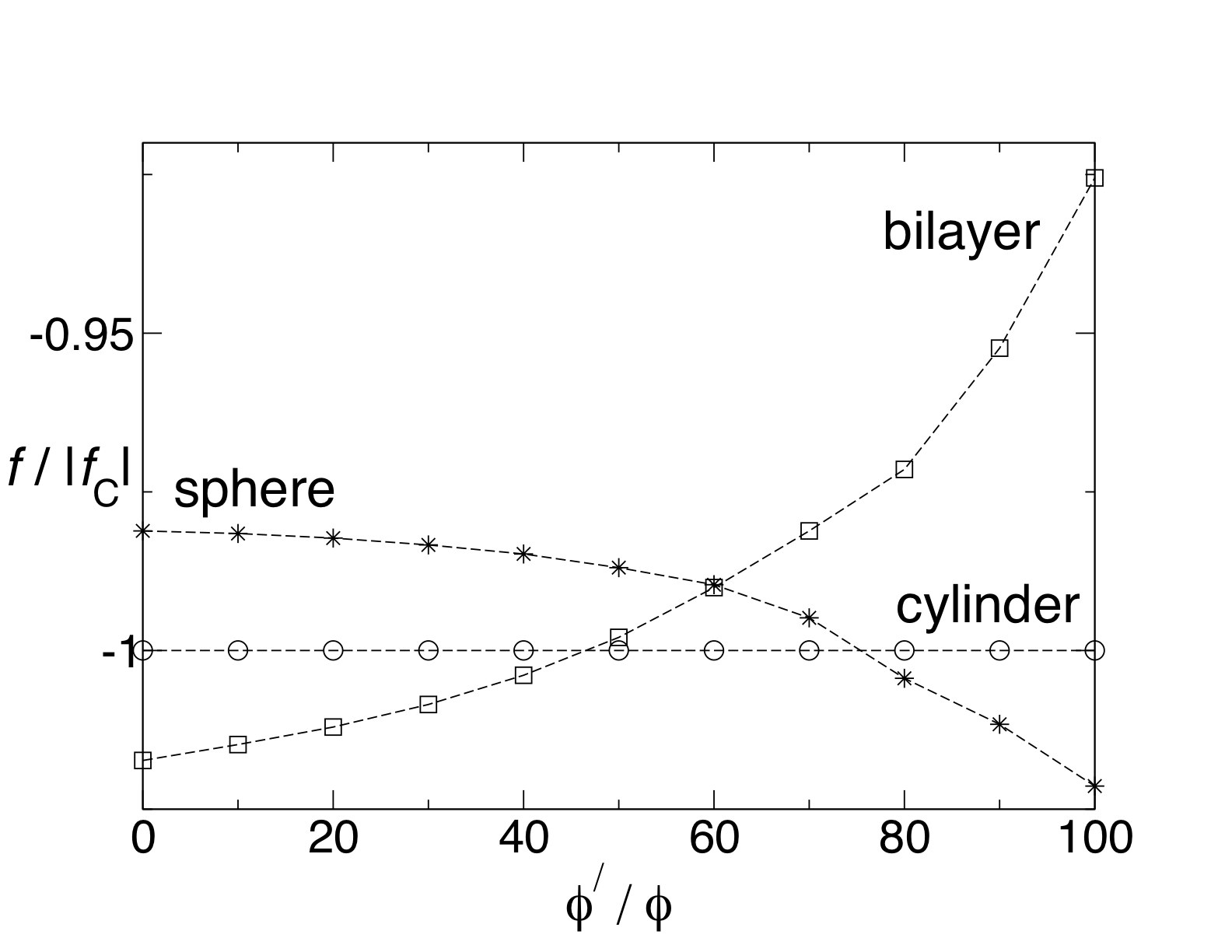}
\caption{\label{ws_fig} Free-energy densities of the
  optimum sphere (stars), cylinder (circles) and bilayer (squares) as
  the volume fraction $\phi'$ of sphere-former is increased at
  fixed overall volume fraction $\phi=10\,\%$. In this
  case, the micelle-forming species is a sphere-former with
  $R^2_0=2a^2N$, and $1/4$ of its monomers are hydrophobic. As $\phi'$ is
  increased, the structure with the lowest free energy changes from
  the bilayer, then to the cylinder and finally to the sphere.}
\end{figure}

We now consider the effect of adding a curvature-stabilizing
amphiphile with a shorter hydrophilic block (but the same size
hydrophobic block) to our system of
symmetric lamella-former. The mean-square end-to-end distance of this
molecule is $R_0^2=2a^2N$ and $1/4$ of its monomers are hydrophobic. The free-energy density plot is seen in Figure \ref{ws_fig}. Despite the significant difference in the length
of the sphere-forming amphiphile from the case considered above, the transitions from sphere to cylinder and
cylinder to bilayer occur at very nearly the same volume fractions as
before. The only notable difference between the two figures is that the
various free energies are closer together, due to
the fact that the two species of amphiphile are closer in length than
before.

\begin{figure}
\includegraphics[width=\linewidth]{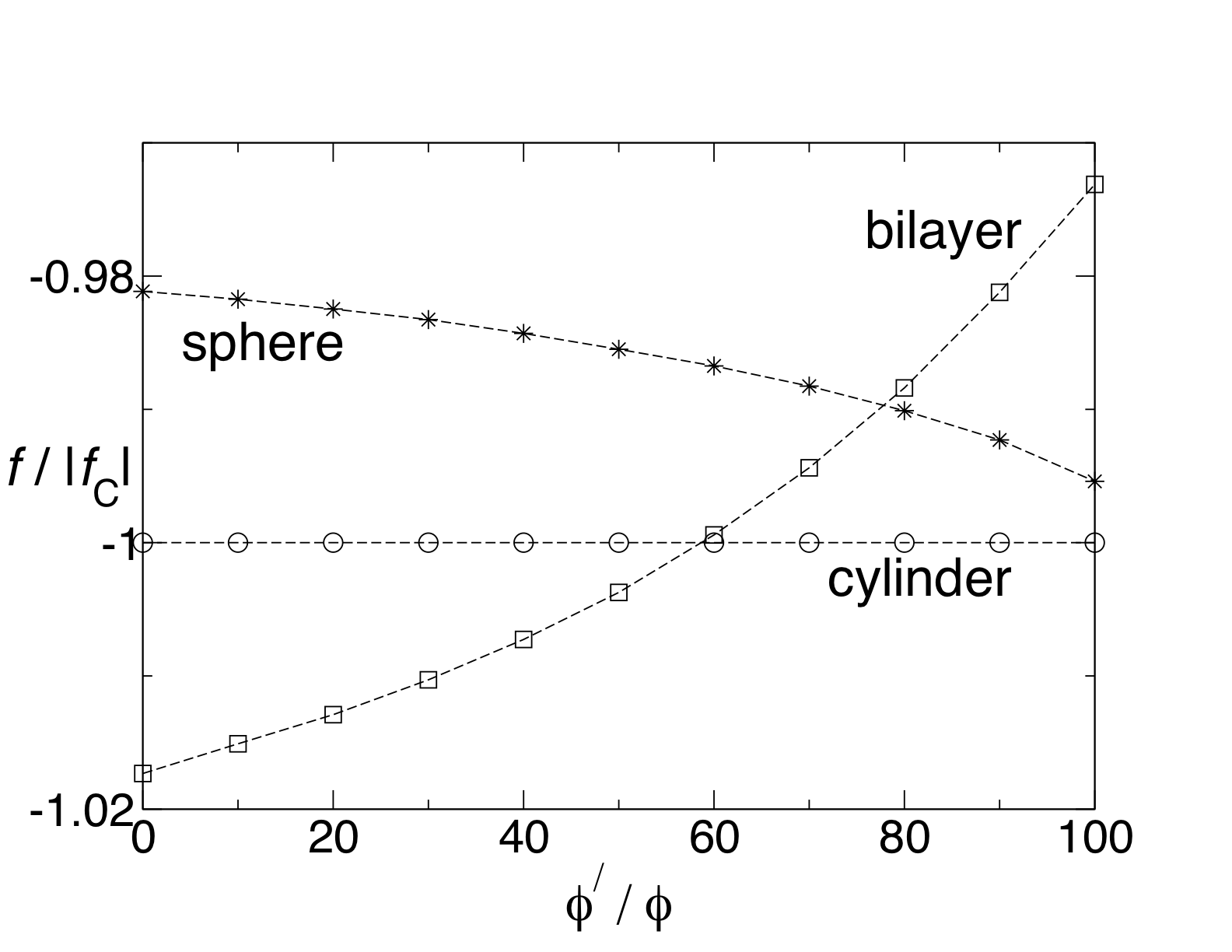}
\caption{\label{wws_fig} Free-energy densities of the
  optimum sphere (stars), cylinder (circles) and bilayer (squares) as
  the volume fraction $\phi'$ of micelle-former is increased at fixed
  overall volume fraction $\phi=10\,\%$. In this case, the micelle-forming
  species is rather short: its mean-square end-to-end distance is $1.5a^2N$, and $1/3$ of
its monomers are hydrophobic. As $\phi'$ is increased, a
  transition from bilayers to cylinders is observed, although the
  cylinder-sphere transition is no longer present. }
\end{figure}

A much more pronounced difference in phenomenology is seen when the
hydrophilic block of the
micelle-forming amphiphile is shortened still further, so that the
mean-squared end-to-end distance of this molecule is $R_0^2=1.5a^2N$ and its fraction of
hydrophobic monomers is $1/3$. From the free-energy plot in Figure \ref{wws_fig}, we see that the lamella-cylinder transition is
shifted to from $\phi'/\phi\approx 50\%$ to $\phi'/\phi\approx 60\%$, and the cylinder-sphere
transition disappears entirely. The implications of this for the
stability of connected structures will be discussed later. In
addition, and following the trend observed above, the
difference between the free energies of the three structures becomes
smaller due to the lesser degree of asymmetry between the two
amphiphiles. 

\begin{figure}
\includegraphics[width=\linewidth]{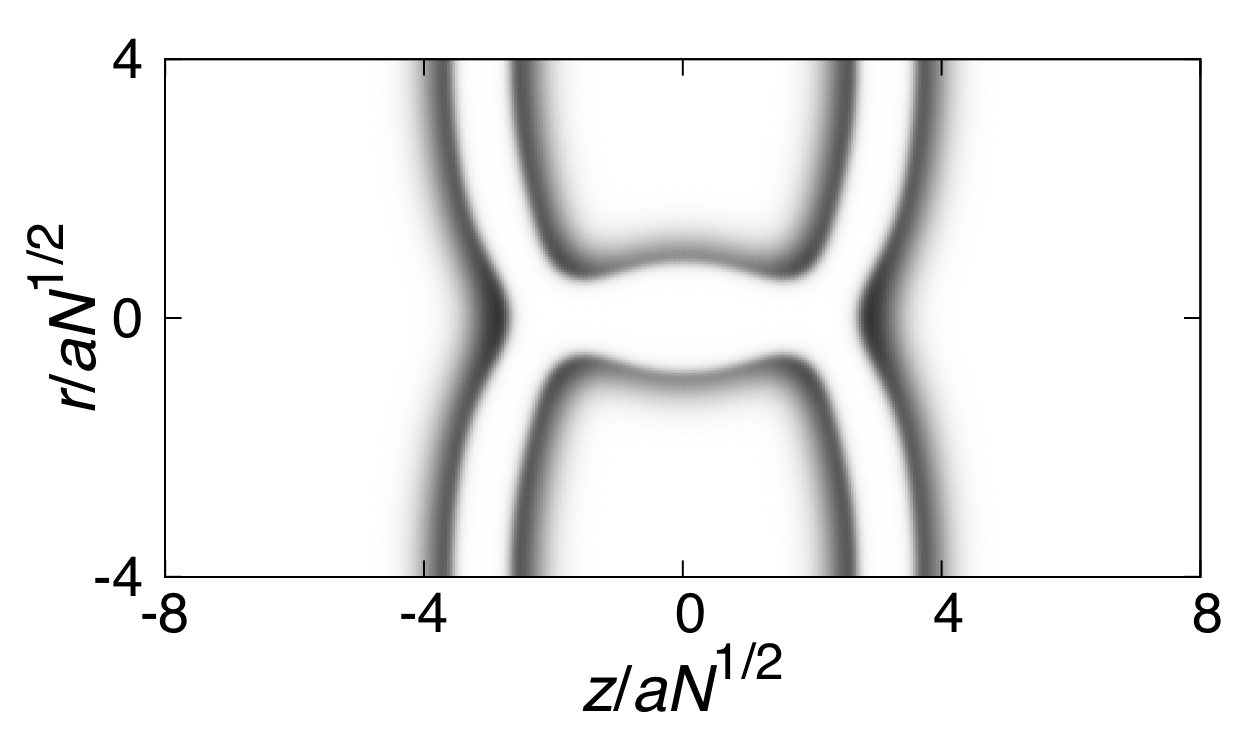}
\caption{\label{A_fig} Density of hydrophilic A-blocks for
a connected bilayer structure in a system of symmetric AB amphiphiles
dissolved in A homopolymer solvent. Dark regions indicate high
hydrophilic block
volume fraction. The plot is in cylindrical polar
coordinates, so that the connecting thread appears as a horizontal structure
along the $z$-axis between the two vertical bilayers.}
\end{figure}

Having established the basic phenomenology of adding
curvature-stabilizing amphiphile to a solution of lamella-former, we
now turn our attention to the structure of particular interest here: a
system of two flat, parallel bilayers connected by a thread-like
cylindrical micelle. As discussed in the methods section, we
consider a cylindrical box of copolymer and work in the canonical
ensemble. To form the connected bilayer, we initiate the SCFT
iteration with a simple ansatz which has the basic form of a connected
bilayer, but no detailed information about the curvature of the
structure or the distribution of the material between its regions. In
the case of a solution of lamella-former with no added
sphere-forming species, we find the structure plotted in
cylindrical polar coordinates in Figure \ref{A_fig}. The plot shows
the density of the hydrophilic A blocks, and the cylindrical thread along the
$z$-axis appears as a horizontal
structure in the centre of the diagram, connecting the two vertical
bilayers. Note also the negative curvature of the surface in the regions
where the thread joins the bilayers. Junctions of this form have also been seen in lattice SCFT calculations by J\'{o}dar-Reyes and Leermakers \cite{jodar-reyes,jodar-reyes2} on the closely-related problem of cylindrical micelles bridging two fixed hydrophilic surfaces.

\begin{figure}
\includegraphics[width=\linewidth]{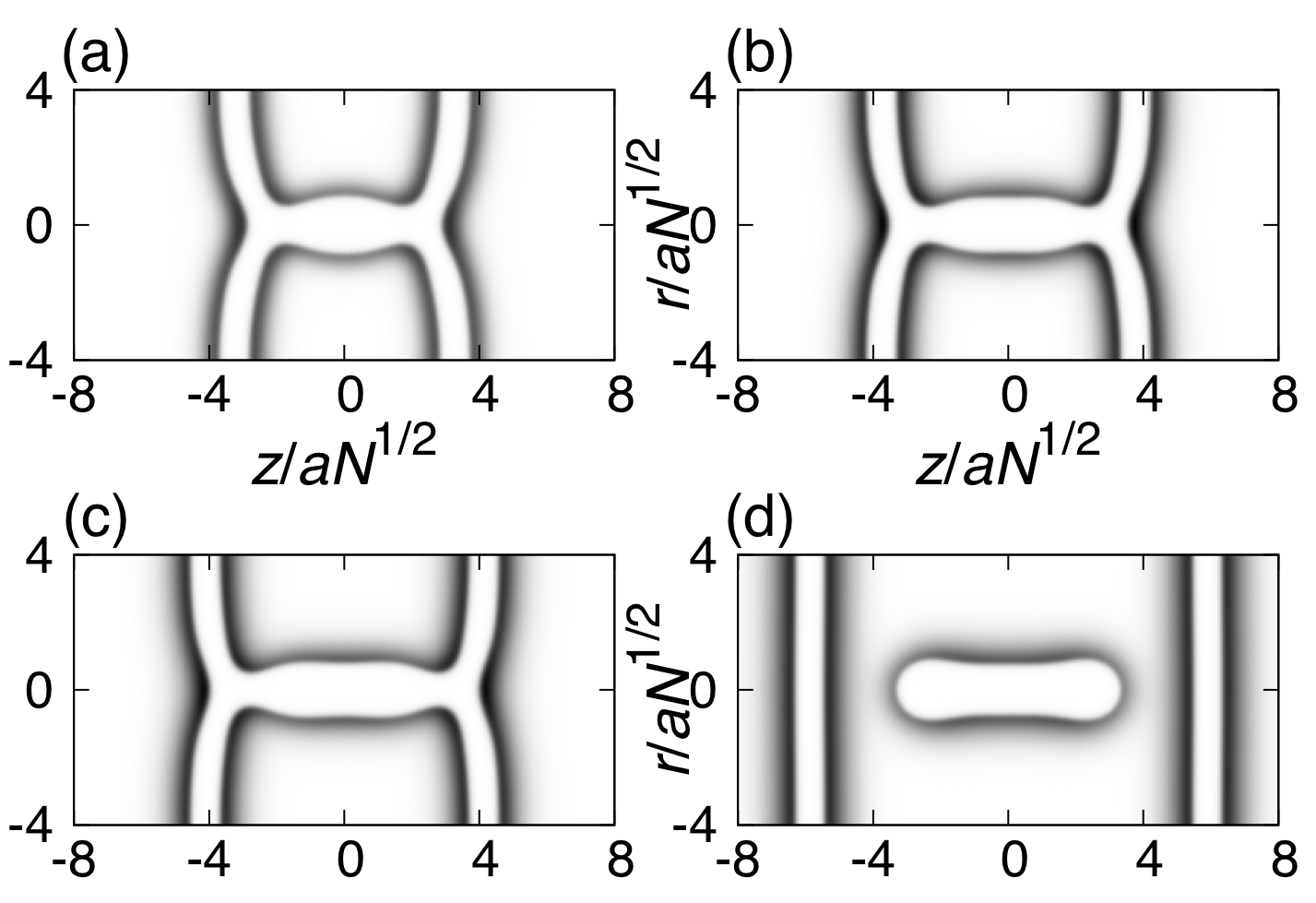}
\caption{\label{A4_fig}  Total density of hydrophilic A-blocks for
connected bilayer structures as lamella-forming polymers are replaced
by sphere-formers, keeping the total number of polymers constant. Cylindrical polar coordinates are used, and dark regions indicate high
hydrophilic block
volume fraction. In (a), $10\,\%$ by volume of all amphiphiles
are sphere-forming. This is increased to $25\,\%$ and then $33\,\%$ in
(b) and (c), resulting in growth of the thread connecting the
two bilayers. In (d), the addition of $50\,\%$ sphere-former
destabilizes the negative curvature regions around the junctions and
leads to a breakdown of the connected structure.}
\end{figure}

To illustrate the effect of adding sphere-forming amphiphile to the
system, we use the following numerical scheme. We fix the number of
amphiphilic molecules in the system. Since the two species of
amphiphile contain the same number of hydrophobic monomers, this
corresponds to fixing the amount of hydrophobic block in the
system. However, we gradually replace the symmetric lamella-forming
diblocks with sphere-forming copolymers. We first focus
on the intermediate-length micelle former with $R_0^2=2a^2N$. The effect of this on the
relative amount of material in the thread and bilayer sections of the
connected structure is shown in Figure \ref{A4_fig}, where we show the
shape of the structures formed by plotting the
total density of hydrophilic A-blocks (lamella former plus sphere former). In Figure \ref{A4_fig} (a), only
$10\,\%$ by volume of all amphiphiles are sphere-forming, and the
structure is little different to that formed in a system of pure
lamella-formers. In Figure \ref{A4_fig} (b), however, we have increased the number
of sphere-formers to $25\,\%$ of all amphiphiles, and a lengthening of the central thread-like region
may be observed. This effect becomes stronger in the plot of Figure \ref{A4_fig}
(c), where $33\,\%$ of the amphiphiles are sphere-forming. We have thus demonstrated how adding sphere-forming species to
the system leads to a transfer of material to the more highly-curved
cylindrical section, and thus that adding such material may make the
formation of thread-like structures more favorable. This is in line
with our simpler one-dimensional calculations discussed above, where
adding sphere-former was shown to lower the free energy of cylindrical
micelles.

However, the existence of connected vesicles such as those seen in the
experiments of Zidovska {\em et al.} \cite{zidovska}. depends not only on the formation
of cylindrical micelles, but also on the stability of the junctions
connecting these to the bilayers. In Figure \ref{A4_fig} (d),
we show the result of replacing $50\,\%$ by volume of the
lamella-formers with sphere-formers. Here, the cylindrical micelle
breaks away from the bilayers that it previously connected, as the
regions of negative curvature associated with the junctions become
unstable in the presence of large amounts of molecules that strongly
favor positive curvature. We note that this solution to the SCFT
equations is found when the initial ansatz has the form of a connected
structure. Despite having varied the system size and the form of the
ansatz, we have been unable to find a connected solution for such high
amounts of sphere-former. This is in agreement with the experiments of
Zidovska {\em et al.} \cite{zidovska}, where connected vesicles are found only in a
narrow range of sphere-former concentrations. Although adding
sphere-former can stabilize thread-like structures, the addition of
too high a quantity of such amphiphiles will destabilize the junctions
connecting these to vesicles. The competition between these two
effects thus restricts the range in which connected vesicle structures
may be formed.

To gain further insight into the mechanism behind the growth and
eventual splitting off of the thread-like structure, we now consider
the distribution of the two amphiphile species within the
aggregates. We quantify this using an {\em enhancement factor}
$\eta(\mathbf{r})$, which we define as
\begin{equation}
\eta(\mathbf{r})=\frac{\phi_\text{B2}(\mathbf{r})}{\phi_\text{B}(\mathbf{r})}\frac{\overline{\phi}_\text{B}}{\overline{\phi}_\text{B2}}
\label{enhance}
\end{equation}
This quantity tells us how much the volume fraction of sphere-former is
enhanced with respect to that of the lamella-former at a given point
in the system. Since $\eta(\mathbf{r})$ is normalized with respect to the total
volume fractions of each species, values greater than one correspond to
enhancement of the sphere-former concentration and values less than
one correspond to depletion.

\begin{figure}
\includegraphics[width=\linewidth]{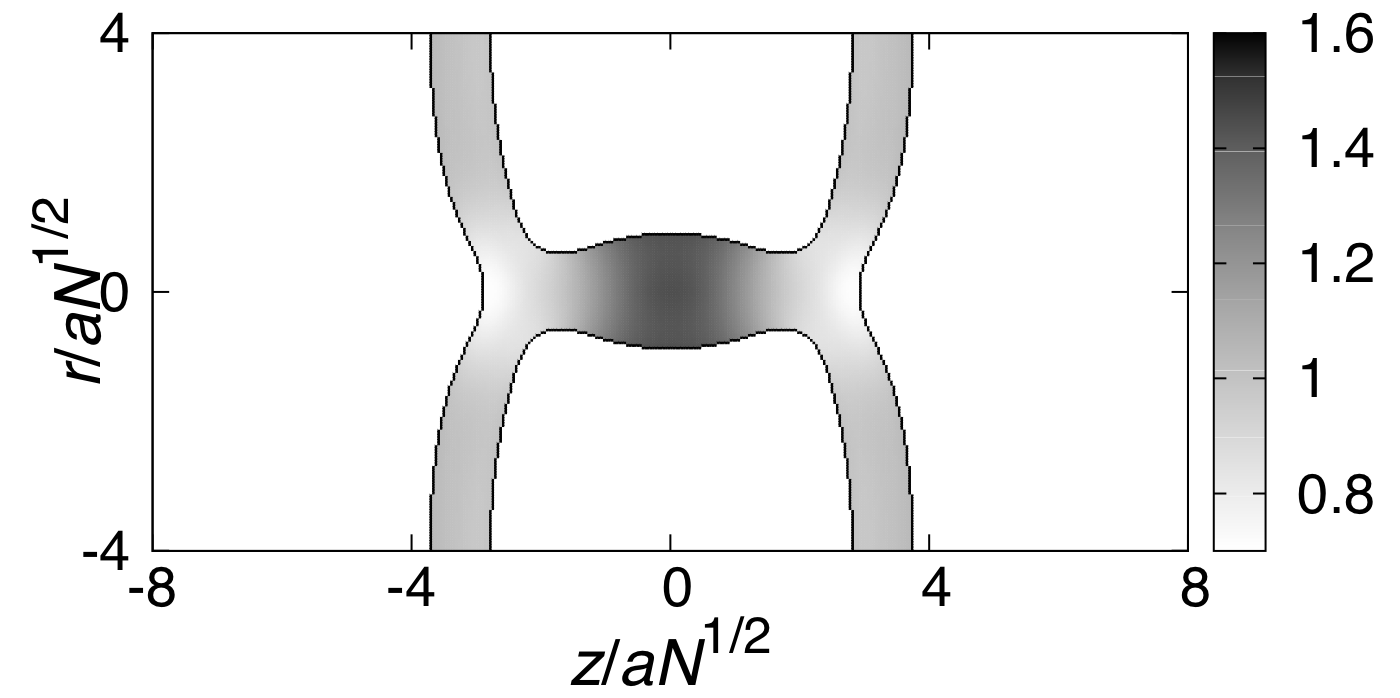}
\caption{\label{eta9_fig} Enhancement factor
  $\eta(\mathbf{r})$ in the hydrophobic core region of a connected
  bilayer formed from $10\,\%$ sphere-formers and $90\,\%$
  lamella-formers. The dark areas ($\eta>1$) show regions where the
  local volume fraction of sphere-former is enhanced, and light
  regions ($\eta<1$) show regions where it is depleted. The
boundary of the core region (defined as the locus of points where the
local volume fractions of hydrophilic and hydrophobic blocks are
equal) is marked with a black line.}
\end{figure}

Figure \ref{eta9_fig} shows a plot of this quantity within the hydrophobic core
region of the connected bilayer formed when $10\,\%$ of the
amphiphiles are sphere-formers (the system shown in Figure \ref{A4_fig} (a)). The segregation of sphere-formers
to the central thread can immediately be seen, as can the depletion of
this species in the negative
curvature regions around the junctions. This result clearly shows the
presence of the significant spontaneous sorting of amphiphiles by
curvature, even at the current simple level of modeling. Furthermore,
it shows that the pictorial model of connected vesicle formation in terms of
lipid segregation put forward by Zidovska {\em et al.} \cite{zidovska} can be
substantiated in explicit calculations.

\begin{figure}
\includegraphics[width=\linewidth]{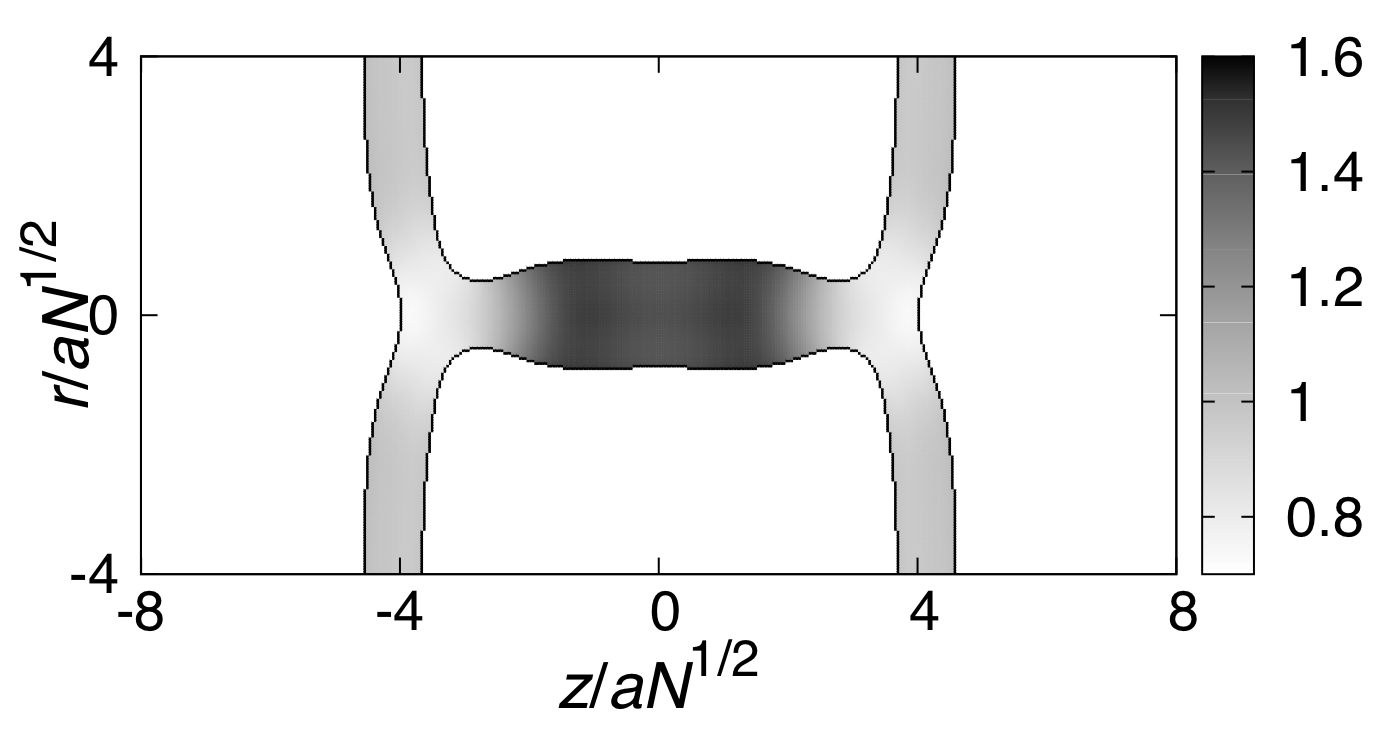}
\caption{\label{eta2_fig} Enhancement factor
  $\eta(\mathbf{r})$ in the hydrophobic core region of a connected
  bilayer formed from $33\,\%$ sphere-formers and $67\,\%$
  lamella-formers. The dark areas ($\eta>1$) show regions where the
  local volume fraction of sphere-former is enhanced, and light
  regions ($\eta<1$) show regions where it is depleted. A hint of the
 underlying cylindrical micelle structure can be seen in
this result: the sphere-formers have begun to cluster into regions at
either end of the thread structure. These clusters will eventually
form the end-caps of the separate cylindrical micelle shown in Figure \ref{A4_fig}
(d).}
\end{figure}

Further insight into the role of amphiphile segregation in aggregate
formation can be gained by plotting the enhancement factor for the
higher sphere-former fraction of $33\,\%$. This is the highest volume
fraction considered before the junctions connecting the bilayers
become unstable. In this case, we see in Figure \ref{eta2_fig} that there is a slightly thinner
region in the center of the thread. On either side of this are two
slightly darker patches, corresponding to higher concentrations of
sphere-former. These features arise from the underlying cylindrical micelle structure of the thread \cite{jodar-reyes}. When the cylindrical micelle breaks away from the bilayers at higher sphere-former concentrations, the clusters of sphere-formers close to the junctions
will form the end-caps of the structure
shown in Figure \ref{A4_fig} (d).

\begin{figure}
\includegraphics[width=\linewidth]{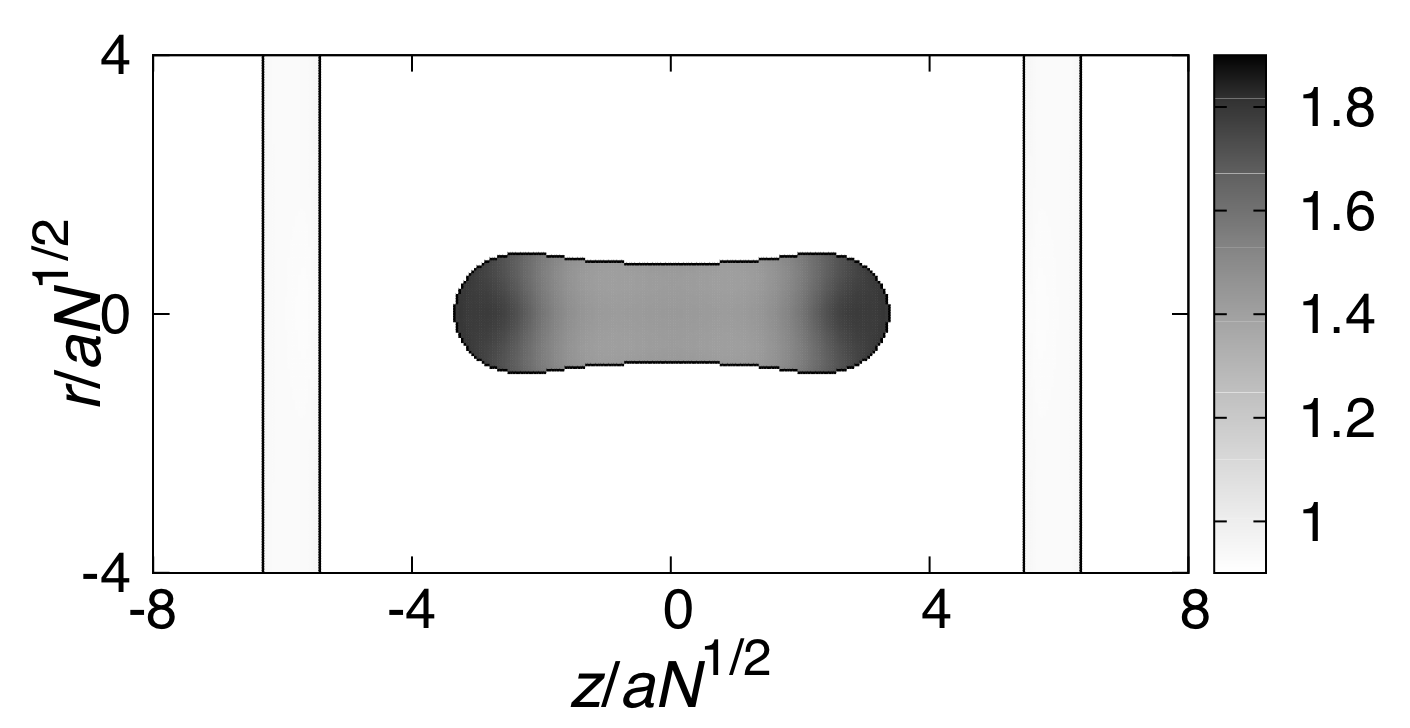}
\caption{\label{eta1_fig} Enhancement factor
  $\eta(\mathbf{r})$ in the hydrophobic core region of a split structure
  (two bilayers and a cylindrical micelle) formed from $50\,\%$ sphere-formers and $50\,\%$
  lamella-formers. The dark areas ($\eta>1$) show regions where the
  local volume fraction of sphere-former is enhanced, and light
  regions ($\eta<1$) show regions where it is depleted. Note the
  much higher concentration of sphere-formers in the cylindrical
  micelle and especially in its end-caps.}
\end{figure}

Next, we demonstrate the segregation of the amphiphiles in this split
structure. The enhancement factor $\eta(\mathbf{r})$ is plotted in Figure \ref{eta1_fig}. We see immediately that there is a sharp
difference in the fraction of sphere-former between the central cylindrical
micelle and the flat bilayers. In addition, we note that the range of
values taken by $\eta(\mathbf{r})$ has changed: for the two lower
sphere-former concentrations shown above, $\eta(\mathbf{r})$ ran from
approximately $0.7$ to $1.6$; at $50\,\%$ sphere-former, it ranges
from around $0.9$ to $1.9$. The reason for the increase in the upper
limit of the range is the existence of the highly-curved end-caps of
the separate cylindrical micelle, which are close to spherical
micelles in structure and thus strongly favored by the
sphere-former. In turn, the disappearance of the regions of negative curvature
at the cylinder-bilayer junctions accounts for the rise in the lower
limit of $\eta(\mathbf{r})$ to a value closer to one.

Finally, we consider the sensitivity of the amphiphile segregation and
junction stability to
the structure of the micelle-forming amphiphile. In addition to the sphere-forming
additive of mean-square end-to-end distance $R_0^2=2a^2N$, we have also
generated analogous results for the shorter and longer curvature
formers introduced earlier, and now present a selection of these.

\begin{figure}
\includegraphics[width=\linewidth]{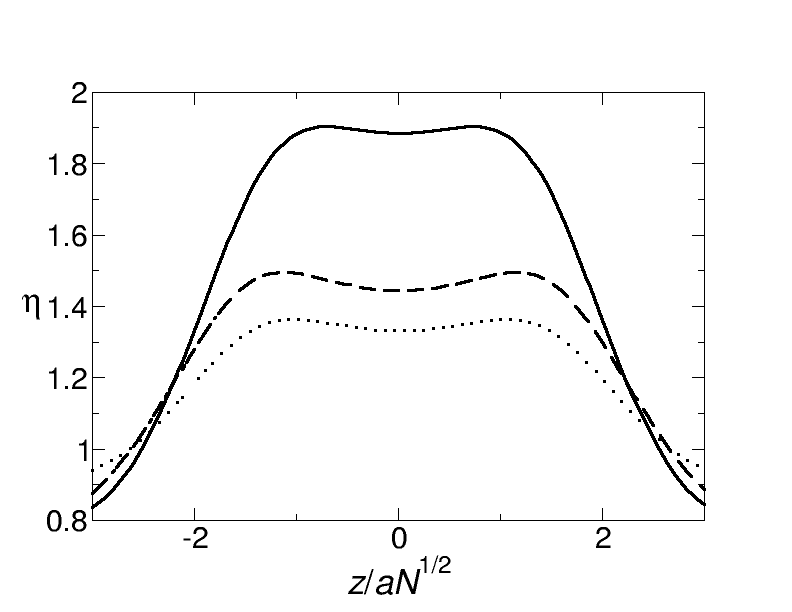}
\caption{\label{etacut_fig} Cuts through the enhancement factor
  $\eta(\mathbf{r})$ along the axis of the cylindrical micelle ($r=0$) for a connected bilayer formed from $33\,\%$ micelle-formers and $67\,\%$
  lamella-formers. The results for the three micelle-formers $R_0^2=1.5a^2N$, $R_0^2=2a^2N$ and $R_0^2=3a^2N$ are shown with dotted, dashed and full lines respectively. }
\end{figure}

In Figure \ref{etacut_fig}, we plot cuts through the enhancement factor
$\eta(\mathbf{r})$ along the axis of the cylindrical micelle ($r=0$) for all three micelle-formers ($R_0^2=1.5a^2N$,$2a^2N$ and $3a^2N$). In each case, we consider a system where $33\,\%$ of the amphiphiles are
micelle-formers, as in Figures \ref{A4_fig} (c) and \ref{eta2_fig} above. We immediately see that the segregation of the micelle-forming species to the thread-like micelle increases sharply as its hydrophilic block length is increased. Indeed, the peak value of $\eta(\mathbf{r})$ increases from around $1.35$ for the shortest micelle-formers ($R_0^2=1.5a^2N$) to just over $1.9$ for the longest ($R_0^2=3a^2N$). We note in particular that changing the mean-square end-to-end distance of the micelle-former from $R_0^2=2a^2N$ to $R_0^2=3a^2N$ has a especially strong effect on the segregation. However, the boundaries of the shape transitions between spheres, cylinders and bilayers (Figures \ref{ms_fig} and \ref{ws_fig}) for these two systems are very similar. It therefore appears that
the amphiphile structure can have a greater effect on the sorting of amphiphiles than on the shape transitions.

Finally, we also note the effect of amphiphile architecture on the
stability of the cylinder-bilayer junctions. In the case of the
stronger sphere-forming additive with $R_0^2=3a^2N$, the junctions
become unstable when approximately $50\,\%$ of amphiphiles are
hydrophobic, as in the case discussed above with $R_0^2=2a^2N$. In contrast, in the case of the shorter amphiphile with
$R_0^2=1.5a^2N$, the connected bilayer remains stable at
this composition. The split structure can however be found by
increasing the percentage of micelle-former to $67\,\%$.

\section{Conclusions}\label{conclusions}

Using a coarse-grained mean-field approach (self-consistent field theory) we have modeled
several aspects of the formation of vesicles connected by thread-like
micelles in amphiphile mixtures. Firstly, we
have performed (effectively one-dimensional) calculations on spherical
micelles, infinite cylinders and infinite bilayers to show that the addition of sphere-forming amphiphile to a solution of
lamella-former can lower the free energy of cylindrical structures, thereby encouraging their formation. Next, through more detailed
calculations on the connected bilayer structure, we have demonstrated that increasing the concentration of sphere-former leads to a growth of
the thread linking the two bilayers. This shows in a direct way how
changing the composition of a solution of amphiphiles may change the shape
of the aggregates that they form, and, in particular change the
relative amounts of regions of different curvature.

Our calculations also yield information on the
distrubution of the two species within the structure, and reveal that the concentration of sphere-former is
significantly higher in the thread. This is in line with the picture
proposed by Zidovska {\em et al.} \cite{zidovska} of the connected vesicles observed in
their experiments. Furthermore, it shows that the phenomenon of amphiphile sorting by
curvature is
present even when the molecules are modeled at a very simple level and
therefore might expect to be observed in a wide variety of systems. We note that amphiphile sorting is observed in our system even though the two lipids are formed of the same two types of monomer A and B and would not demix in a structure of uniform curvature such as a flat bilayer. This is in contrast to recent experiments by Sorre {\em et al.} \cite{sorre}, where proximity to a demixing point is found to be essential for lipid sorting to occur. There are two reason for this. Firstly, we consider amphiphiles that individually form different structures (micelles and bilayers respectively) and so segregate strongly to regions of different curvature within the connected bilayer structure. The amphiphiles used by Sorre {\em et al.}, however, both form bilayers \cite{shipley,ulrich} (albeit of different bending rigidities). The extra factor of demixing is thus required for lipid sorting to occur in their system. Secondly, we study bilayers connected by thread-like micelles rather than hollow bilayer tubes. The difference in curvature between the two regions of our structures is thus very large, and a clear segregation of the two species is seen even in the absence of phase separation.

We then investigate the stability of the junctions between the thread
and the bilayers, and show that adding larger amounts of sphere-former
can cause the connected structures to break down as the regions of
negative curvature around the junctions become more unfavorable. Both this effect and
the degree of segregation of the amphiphiles are shown to be rather
sensitive to the length of the hydrophilic block of the micelle-forming
amphiphiles. In particular, the degree of segregation (or efficiency
of sorting) is shown to have a particularly strong dependence on the hydrophilic block
length of the micelle-former.


\begin{thebibliography}{38}
\expandafter\ifx\csname natexlab\endcsname\relax\def\natexlab#1{#1}\fi
\expandafter\ifx\csname bibnamefont\endcsname\relax
  \def\bibnamefont#1{#1}\fi
\expandafter\ifx\csname bibfnamefont\endcsname\relax
  \def\bibfnamefont#1{#1}\fi
\expandafter\ifx\csname citenamefont\endcsname\relax
  \def\citenamefont#1{#1}\fi
\expandafter\ifx\csname url\endcsname\relax
  \def\url#1{\texttt{#1}}\fi
\expandafter\ifx\csname urlprefix\endcsname\relax\def\urlprefix{URL }\fi
\providecommand{\bibinfo}[2]{#2}
\providecommand{\eprint}[2][]{\url{#2}}

\bibitem[{\citenamefont{Jain and Bates}(2003)}]{jain_bates}
\bibinfo{author}{\bibfnamefont{S.}~\bibnamefont{Jain}} \bibnamefont{and}
  \bibinfo{author}{\bibfnamefont{F.~S.} \bibnamefont{Bates}},
  \bibinfo{journal}{Science} \textbf{\bibinfo{volume}{300}},
  \bibinfo{pages}{460} (\bibinfo{year}{2003}).

\bibitem[{\citenamefont{Battaglia and Ryan}(2006)}]{battaglia_ryan}
\bibinfo{author}{\bibfnamefont{G.}~\bibnamefont{Battaglia}} \bibnamefont{and}
  \bibinfo{author}{\bibfnamefont{A.~J.} \bibnamefont{Ryan}},
  \bibinfo{journal}{Journal of Physical Chemistry B}
  \textbf{\bibinfo{volume}{110}}, \bibinfo{pages}{10272}
  (\bibinfo{year}{2006}).

\bibitem[{\citenamefont{Smart et~al.}(2010)\citenamefont{Smart, Ryan, Howse,
  and Battaglia}}]{smart}
\bibinfo{author}{\bibfnamefont{T.~P.} \bibnamefont{Smart}},
  \bibinfo{author}{\bibfnamefont{A.~J.} \bibnamefont{Ryan}},
  \bibinfo{author}{\bibfnamefont{J.~R.} \bibnamefont{Howse}}, \bibnamefont{and}
  \bibinfo{author}{\bibfnamefont{G.}~\bibnamefont{Battaglia}},
  \bibinfo{journal}{Langmuir} \textbf{\bibinfo{volume}{26}},
  \bibinfo{pages}{7425} (\bibinfo{year}{2010}).

\bibitem[{\citenamefont{Howse et~al.}(2009)\citenamefont{Howse, Jones,
  Battaglia, Ducker, Leggett, and Ryan}}]{howse}
\bibinfo{author}{\bibfnamefont{J.~R.} \bibnamefont{Howse}},
  \bibinfo{author}{\bibfnamefont{R.~A.~L.} \bibnamefont{Jones}},
  \bibinfo{author}{\bibfnamefont{G.}~\bibnamefont{Battaglia}},
  \bibinfo{author}{\bibfnamefont{R.~E.} \bibnamefont{Ducker}},
  \bibinfo{author}{\bibfnamefont{G.~J.} \bibnamefont{Leggett}},
  \bibnamefont{and} \bibinfo{author}{\bibfnamefont{A.~J.} \bibnamefont{Ryan}},
  \bibinfo{journal}{Nature Materials} \textbf{\bibinfo{volume}{8}},
  \bibinfo{pages}{507} (\bibinfo{year}{2009}).

\bibitem[{\citenamefont{Kim et~al.}(2005)\citenamefont{Kim, Dalhaimer,
  Christian, and Discher}}]{kim}
\bibinfo{author}{\bibfnamefont{Y.}~\bibnamefont{Kim}},
  \bibinfo{author}{\bibfnamefont{P.}~\bibnamefont{Dalhaimer}},
  \bibinfo{author}{\bibfnamefont{D.~A.} \bibnamefont{Christian}},
  \bibnamefont{and} \bibinfo{author}{\bibfnamefont{D.~E.}
  \bibnamefont{Discher}}, \bibinfo{journal}{Nanotechnology}
  \textbf{\bibinfo{volume}{16}}, \bibinfo{pages}{S484} (\bibinfo{year}{2005}).

\bibitem[{\citenamefont{Lomas et~al.}(2007)\citenamefont{Lomas, Canton,
  MacNeil, Du, Armes, Ryan, Lewis, and Battaglia}}]{lomas}
\bibinfo{author}{\bibfnamefont{H.}~\bibnamefont{Lomas}},
  \bibinfo{author}{\bibfnamefont{I.}~\bibnamefont{Canton}},
  \bibinfo{author}{\bibfnamefont{S.}~\bibnamefont{MacNeil}},
  \bibinfo{author}{\bibfnamefont{J.}~\bibnamefont{Du}},
  \bibinfo{author}{\bibfnamefont{S.~P.} \bibnamefont{Armes}},
  \bibinfo{author}{\bibfnamefont{A.~J.} \bibnamefont{Ryan}},
  \bibinfo{author}{\bibfnamefont{A.~L.} \bibnamefont{Lewis}}, \bibnamefont{and}
  \bibinfo{author}{\bibfnamefont{G.}~\bibnamefont{Battaglia}},
  \bibinfo{journal}{Advanced Materials} \textbf{\bibinfo{volume}{19}},
  \bibinfo{pages}{4238} (\bibinfo{year}{2007}).

\bibitem[{\citenamefont{Kinning et~al.}(1988)\citenamefont{Kinning, Winey, and
  Thomas}}]{kinning_winey_thomas}
\bibinfo{author}{\bibfnamefont{D.~J.} \bibnamefont{Kinning}},
  \bibinfo{author}{\bibfnamefont{K.~I.} \bibnamefont{Winey}}, \bibnamefont{and}
  \bibinfo{author}{\bibfnamefont{E.~L.} \bibnamefont{Thomas}},
  \bibinfo{journal}{Macromolecules} \textbf{\bibinfo{volume}{21}},
  \bibinfo{pages}{3502} (\bibinfo{year}{1988}).

\bibitem[{\citenamefont{Schuetz et~al.}(2011)\citenamefont{Schuetz, Greenall,
  Bent, Furzeland, Atkins, Butler, McLeish, and Buzza}}]{schuetz}
\bibinfo{author}{\bibfnamefont{P.}~\bibnamefont{Schuetz}},
  \bibinfo{author}{\bibfnamefont{M.~J.} \bibnamefont{Greenall}},
  \bibinfo{author}{\bibfnamefont{J.}~\bibnamefont{Bent}},
  \bibinfo{author}{\bibfnamefont{S.}~\bibnamefont{Furzeland}},
  \bibinfo{author}{\bibfnamefont{D.}~\bibnamefont{Atkins}},
  \bibinfo{author}{\bibfnamefont{M.~F.} \bibnamefont{Butler}},
  \bibinfo{author}{\bibfnamefont{T.~C.~B.} \bibnamefont{McLeish}},
  \bibnamefont{and} \bibinfo{author}{\bibfnamefont{D.~M.~A.}
  \bibnamefont{Buzza}}, \bibinfo{journal}{to appear in Soft Matter}
  (\bibinfo{year}{2011}).

\bibitem[{\citenamefont{Sorre et~al.}(2009)\citenamefont{Sorre, Callan-Jones,
  Manneville, Nassoy, Joanny, Prost, Goud, and Bassereau}}]{sorre}
\bibinfo{author}{\bibfnamefont{B.}~\bibnamefont{Sorre}},
  \bibinfo{author}{\bibfnamefont{A.}~\bibnamefont{Callan-Jones}},
  \bibinfo{author}{\bibfnamefont{J.~B.} \bibnamefont{Manneville}},
  \bibinfo{author}{\bibfnamefont{P.}~\bibnamefont{Nassoy}},
  \bibinfo{author}{\bibfnamefont{J.~F.} \bibnamefont{Joanny}},
  \bibinfo{author}{\bibfnamefont{J.}~\bibnamefont{Prost}},
  \bibinfo{author}{\bibfnamefont{B.}~\bibnamefont{Goud}}, \bibnamefont{and}
  \bibinfo{author}{\bibfnamefont{P.}~\bibnamefont{Bassereau}},
  \bibinfo{journal}{Proceedings of the National Academy of Sciences of the
  United States of America} \textbf{\bibinfo{volume}{106}},
  \bibinfo{pages}{5622} (\bibinfo{year}{2009}).

\bibitem[{\citenamefont{Akiyoshi et~al.}(2003)\citenamefont{Akiyoshi, Itaya,
  Nomura, Ono, and Yoshikawa}}]{akiyoshi}
\bibinfo{author}{\bibfnamefont{K.}~\bibnamefont{Akiyoshi}},
  \bibinfo{author}{\bibfnamefont{A.}~\bibnamefont{Itaya}},
  \bibinfo{author}{\bibfnamefont{S.~M.} \bibnamefont{Nomura}},
  \bibinfo{author}{\bibfnamefont{N.}~\bibnamefont{Ono}}, \bibnamefont{and}
  \bibinfo{author}{\bibfnamefont{K.}~\bibnamefont{Yoshikawa}},
  \bibinfo{journal}{FEBS Letters} \textbf{\bibinfo{volume}{534}},
  \bibinfo{pages}{33} (\bibinfo{year}{2003}).

\bibitem[{\citenamefont{Heinrich et~al.}(2010)\citenamefont{Heinrich, Tian,
  Esposito, and Baumgart}}]{heinrich}
\bibinfo{author}{\bibfnamefont{M.}~\bibnamefont{Heinrich}},
  \bibinfo{author}{\bibfnamefont{A.}~\bibnamefont{Tian}},
  \bibinfo{author}{\bibfnamefont{C.}~\bibnamefont{Esposito}}, \bibnamefont{and}
  \bibinfo{author}{\bibfnamefont{T.}~\bibnamefont{Baumgart}},
  \bibinfo{journal}{Proceedings of the National Academy of Sciences of the
  United States of America} \textbf{\bibinfo{volume}{107}},
  \bibinfo{pages}{7208} (\bibinfo{year}{2010}).

\bibitem[{\citenamefont{Zidovska et~al.}(2009)\citenamefont{Zidovska, Ewert,
  Quispe, Carragher, Potter, and Safinya}}]{zidovska}
\bibinfo{author}{\bibfnamefont{A.}~\bibnamefont{Zidovska}},
  \bibinfo{author}{\bibfnamefont{K.~K.} \bibnamefont{Ewert}},
  \bibinfo{author}{\bibfnamefont{J.}~\bibnamefont{Quispe}},
  \bibinfo{author}{\bibfnamefont{B.}~\bibnamefont{Carragher}},
  \bibinfo{author}{\bibfnamefont{C.~S.} \bibnamefont{Potter}},
  \bibnamefont{and} \bibinfo{author}{\bibfnamefont{C.~R.}
  \bibnamefont{Safinya}}, \bibinfo{journal}{Langmuir}
  \textbf{\bibinfo{volume}{25}}, \bibinfo{pages}{2979} (\bibinfo{year}{2009}).

\bibitem[{\citenamefont{Edwards}(1965)}]{edwards}
\bibinfo{author}{\bibfnamefont{S.~F.} \bibnamefont{Edwards}},
  \bibinfo{journal}{Proc.\ Phys.\ Soc.} \textbf{\bibinfo{volume}{85}},
  \bibinfo{pages}{613} (\bibinfo{year}{1965}).

\bibitem[{\citenamefont{Maniadis et~al.}(2007)\citenamefont{Maniadis, Lookman,
  Kober, and Rasmussen}}]{maniadis}
\bibinfo{author}{\bibfnamefont{P.}~\bibnamefont{Maniadis}},
  \bibinfo{author}{\bibfnamefont{T.}~\bibnamefont{Lookman}},
  \bibinfo{author}{\bibfnamefont{E.~M.} \bibnamefont{Kober}}, \bibnamefont{and}
  \bibinfo{author}{\bibfnamefont{K.~O.} \bibnamefont{Rasmussen}},
  \bibinfo{journal}{Physical Review Letters} \textbf{\bibinfo{volume}{99}},
  \bibinfo{pages}{048302} (\bibinfo{year}{2007}).

\bibitem[{\citenamefont{Drolet and Fredrickson}(1999)}]{drolet_fredrickson}
\bibinfo{author}{\bibfnamefont{F.}~\bibnamefont{Drolet}} \bibnamefont{and}
  \bibinfo{author}{\bibfnamefont{G.~H.} \bibnamefont{Fredrickson}},
  \bibinfo{journal}{Phys.\ Rev.\ Lett.} \textbf{\bibinfo{volume}{83}},
  \bibinfo{pages}{4317} (\bibinfo{year}{1999}).

\bibitem[{\citenamefont{Matsen}(2006)}]{matsen_book}
\bibinfo{author}{\bibfnamefont{M.~W.} \bibnamefont{Matsen}},
  \emph{\bibinfo{title}{Soft Matter}} (\bibinfo{publisher}{Wiley-VCH},
  \bibinfo{address}{Weinheim}, \bibinfo{year}{2006}),
  chap.~\bibinfo{chapter}{2}.

\bibitem[{\citenamefont{Duque}(2003)}]{duque}
\bibinfo{author}{\bibfnamefont{D.}~\bibnamefont{Duque}}, \bibinfo{journal}{J.
  Chem.\ Phys.} \textbf{\bibinfo{volume}{119}}, \bibinfo{pages}{5701}
  (\bibinfo{year}{2003}).

\bibitem[{\citenamefont{Katsov et~al.}(2004)\citenamefont{Katsov, M\"{u}ller,
  and Schick}}]{katsov1}
\bibinfo{author}{\bibfnamefont{K.}~\bibnamefont{Katsov}},
  \bibinfo{author}{\bibfnamefont{M.}~\bibnamefont{M\"{u}ller}},
  \bibnamefont{and} \bibinfo{author}{\bibfnamefont{M.}~\bibnamefont{Schick}},
  \bibinfo{journal}{Biophysical Journal} \textbf{\bibinfo{volume}{87}},
  \bibinfo{pages}{3277} (\bibinfo{year}{2004}).

\bibitem[{\citenamefont{Cavallo et~al.}(2006)\citenamefont{Cavallo, M\"{u}ller,
  and Binder}}]{cavallo}
\bibinfo{author}{\bibfnamefont{A.}~\bibnamefont{Cavallo}},
  \bibinfo{author}{\bibfnamefont{M.}~\bibnamefont{M\"{u}ller}},
  \bibnamefont{and} \bibinfo{author}{\bibfnamefont{K.}~\bibnamefont{Binder}},
  \bibinfo{journal}{Macromolecules} \textbf{\bibinfo{volume}{39}},
  \bibinfo{pages}{9539} (\bibinfo{year}{2006}).

\bibitem[{\citenamefont{Wijmans and Linse}(1995)}]{wijmans_linse}
\bibinfo{author}{\bibfnamefont{C.~M.} \bibnamefont{Wijmans}} \bibnamefont{and}
  \bibinfo{author}{\bibfnamefont{P.}~\bibnamefont{Linse}},
  \bibinfo{journal}{Langmuir} \textbf{\bibinfo{volume}{11}},
  \bibinfo{pages}{3748} (\bibinfo{year}{1995}).

\bibitem[{\citenamefont{Leermakers and
  Scheutjens}(1990)}]{leermakers_scheutjens-shape}
\bibinfo{author}{\bibfnamefont{F.~A.~M.} \bibnamefont{Leermakers}}
  \bibnamefont{and} \bibinfo{author}{\bibfnamefont{J.~M. H.~M.}
  \bibnamefont{Scheutjens}}, \bibinfo{journal}{Journal of Colloid and Interface
  Science} \textbf{\bibinfo{volume}{136}}, \bibinfo{pages}{231}
  (\bibinfo{year}{1990}).

\bibitem[{\citenamefont{Schmid}(1998)}]{schmid_scf_rev}
\bibinfo{author}{\bibfnamefont{F.}~\bibnamefont{Schmid}}, \bibinfo{journal}{J.
  Phys.: Condens.\ Matter} \textbf{\bibinfo{volume}{10}}, \bibinfo{pages}{8105}
  (\bibinfo{year}{1998}).

\bibitem[{\citenamefont{Jones}(2002)}]{jones_book}
\bibinfo{author}{\bibfnamefont{R.~A.~L.} \bibnamefont{Jones}},
  \emph{\bibinfo{title}{Soft Condensed Matter}} (\bibinfo{publisher}{Oxford
  University Press}, \bibinfo{address}{Oxford}, \bibinfo{year}{2002}).

\bibitem[{\citenamefont{M\"{u}ller and Gompper}(2002)}]{mueller}
\bibinfo{author}{\bibfnamefont{M.}~\bibnamefont{M\"{u}ller}} \bibnamefont{and}
  \bibinfo{author}{\bibfnamefont{G.}~\bibnamefont{Gompper}},
  \bibinfo{journal}{Phys.\ Rev.\ E} \textbf{\bibinfo{volume}{66}},
  \bibinfo{pages}{041805} (\bibinfo{year}{2002}).

\bibitem[{\citenamefont{Denesyuk and Gompper}(2006)}]{denesyuk}
\bibinfo{author}{\bibfnamefont{N.~A.} \bibnamefont{Denesyuk}} \bibnamefont{and}
  \bibinfo{author}{\bibfnamefont{G.}~\bibnamefont{Gompper}},
  \bibinfo{journal}{Macromolecules} \textbf{\bibinfo{volume}{39}},
  \bibinfo{pages}{5497} (\bibinfo{year}{2006}).

\bibitem[{\citenamefont{Fredrickson}(2006)}]{fredrickson_book}
\bibinfo{author}{\bibfnamefont{G.~H.} \bibnamefont{Fredrickson}},
  \emph{\bibinfo{title}{The Equilibrium Theory of Inhomogeneous Polymers}}
  (\bibinfo{publisher}{Oxford University Press}, \bibinfo{address}{Oxford},
  \bibinfo{year}{2006}).

\bibitem[{\citenamefont{Matsen}(2004)}]{matsen2004}
\bibinfo{author}{\bibfnamefont{M.~W.} \bibnamefont{Matsen}},
  \bibinfo{journal}{J. Chem.\ Phys.} \textbf{\bibinfo{volume}{121}},
  \bibinfo{pages}{1938} (\bibinfo{year}{2004}).

\bibitem[{\citenamefont{Press et~al.}(1992)\citenamefont{Press, Flannery,
  Teukolsky, and Vetterling}}]{num_rec}
\bibinfo{author}{\bibfnamefont{W.~H.} \bibnamefont{Press}},
  \bibinfo{author}{\bibfnamefont{B.~P.} \bibnamefont{Flannery}},
  \bibinfo{author}{\bibfnamefont{S.~A.} \bibnamefont{Teukolsky}},
  \bibnamefont{and} \bibinfo{author}{\bibfnamefont{W.~T.}
  \bibnamefont{Vetterling}}, \emph{\bibinfo{title}{Numerical Recipes in C}}
  (\bibinfo{publisher}{Cambridge University Press},
  \bibinfo{address}{Cambridge}, \bibinfo{year}{1992}), \bibinfo{edition}{2nd}
  ed.

\bibitem[{\citenamefont{Greenall
  et~al.}(2009{\natexlab{a}})\citenamefont{Greenall, Buzza, and
  McLeish}}]{gbm_macro}
\bibinfo{author}{\bibfnamefont{M.~J.} \bibnamefont{Greenall}},
  \bibinfo{author}{\bibfnamefont{D.~M.~A.} \bibnamefont{Buzza}},
  \bibnamefont{and} \bibinfo{author}{\bibfnamefont{T.~C.~B.}
  \bibnamefont{McLeish}}, \bibinfo{journal}{Macromolecules}
  \textbf{\bibinfo{volume}{42}}, \bibinfo{pages}{5873}
  (\bibinfo{year}{2009}{\natexlab{a}}).

\bibitem[{\citenamefont{Greenall
  et~al.}(2009{\natexlab{b}})\citenamefont{Greenall, Buzza, and
  McLeish}}]{gbm_jcp}
\bibinfo{author}{\bibfnamefont{M.~J.} \bibnamefont{Greenall}},
  \bibinfo{author}{\bibfnamefont{D.~M.~A.} \bibnamefont{Buzza}},
  \bibnamefont{and} \bibinfo{author}{\bibfnamefont{T.~C.~B.}
  \bibnamefont{McLeish}}, \bibinfo{journal}{J. Chem.\ Phys.}
  \textbf{\bibinfo{volume}{131}}, \bibinfo{pages}{034904}
  (\bibinfo{year}{2009}{\natexlab{b}}).

\bibitem[{\citenamefont{Vinson et~al.}(1989)\citenamefont{Vinson, Talmon, and
  Walter}}]{vinson}
\bibinfo{author}{\bibfnamefont{P.~K.} \bibnamefont{Vinson}},
  \bibinfo{author}{\bibfnamefont{Y.}~\bibnamefont{Talmon}}, \bibnamefont{and}
  \bibinfo{author}{\bibfnamefont{A.}~\bibnamefont{Walter}},
  \bibinfo{journal}{Biophysical Journal} \textbf{\bibinfo{volume}{56}},
  \bibinfo{pages}{669} (\bibinfo{year}{1989}).

\bibitem[{\citenamefont{Oberdisse et~al.}(1998)\citenamefont{Oberdisse, Regev,
  and Porte}}]{oberdisse}
\bibinfo{author}{\bibfnamefont{J.}~\bibnamefont{Oberdisse}},
  \bibinfo{author}{\bibfnamefont{O.}~\bibnamefont{Regev}}, \bibnamefont{and}
  \bibinfo{author}{\bibfnamefont{G.}~\bibnamefont{Porte}}, \bibinfo{journal}{J.
  Phys.\ Chem.\ B} \textbf{\bibinfo{volume}{102}}, \bibinfo{pages}{1102}
  (\bibinfo{year}{1998}).

\bibitem[{\citenamefont{Li et~al.}(2009)\citenamefont{Li, Marcelis, Sudholter,
  Stuart, and Leermakers}}]{li}
\bibinfo{author}{\bibfnamefont{F.}~\bibnamefont{Li}},
  \bibinfo{author}{\bibfnamefont{A.~T.~M.} \bibnamefont{Marcelis}},
  \bibinfo{author}{\bibfnamefont{E.~J.~R.} \bibnamefont{Sudholter}},
  \bibinfo{author}{\bibfnamefont{M.~A.~C.} \bibnamefont{Stuart}},
  \bibnamefont{and} \bibinfo{author}{\bibfnamefont{F.~A.~M.}
  \bibnamefont{Leermakers}}, \bibinfo{journal}{Soft Matter}
  \textbf{\bibinfo{volume}{5}}, \bibinfo{pages}{4173} (\bibinfo{year}{2009}).

\bibitem[{\citenamefont{Andelman et~al.}(1994)\citenamefont{Andelman, Kozlov,
  and Helfrich}}]{andelman}
\bibinfo{author}{\bibfnamefont{D.}~\bibnamefont{Andelman}},
  \bibinfo{author}{\bibfnamefont{M.~M.} \bibnamefont{Kozlov}},
  \bibnamefont{and} \bibinfo{author}{\bibfnamefont{W.}~\bibnamefont{Helfrich}},
  \bibinfo{journal}{Europhysics Letters} \textbf{\bibinfo{volume}{25}},
  \bibinfo{pages}{231} (\bibinfo{year}{1994}).

\bibitem[{\citenamefont{J\'{o}dar-Reyes and Leermakers}(2006)}]{jodar-reyes}
\bibinfo{author}{\bibfnamefont{A.~B.} \bibnamefont{J\'{o}dar-Reyes}}
  \bibnamefont{and} \bibinfo{author}{\bibfnamefont{F.~A.~M.}
  \bibnamefont{Leermakers}}, \bibinfo{journal}{J. Phys.\ Chem.\ B}
  \textbf{\bibinfo{volume}{110}}, \bibinfo{pages}{18415}
  (\bibinfo{year}{2006}).

\bibitem[{\citenamefont{J\'{o}dar-Reyes and Leermakers}(2009)}]{jodar-reyes2}
\bibinfo{author}{\bibfnamefont{A.~B.} \bibnamefont{J\'{o}dar-Reyes}}
  \bibnamefont{and} \bibinfo{author}{\bibfnamefont{F.~A.~M.}
  \bibnamefont{Leermakers}}, \bibinfo{journal}{J. Phys.\ Chem.\ B}
  \textbf{\bibinfo{volume}{113}}, \bibinfo{pages}{11186}
  (\bibinfo{year}{2009}).

\bibitem[{\citenamefont{Shipley et~al.}(1974)\citenamefont{Shipley, Avecilla,
  and Small}}]{shipley}
\bibinfo{author}{\bibfnamefont{G.~G.} \bibnamefont{Shipley}},
  \bibinfo{author}{\bibfnamefont{L.~S.} \bibnamefont{Avecilla}},
  \bibnamefont{and} \bibinfo{author}{\bibfnamefont{D.~M.} \bibnamefont{Small}},
  \bibinfo{journal}{Journal of Lipid Research} \textbf{\bibinfo{volume}{15}},
  \bibinfo{pages}{124} (\bibinfo{year}{1974}).

\bibitem[{\citenamefont{Ulrich et~al.}(1994)\citenamefont{Ulrich, Samia, and
  Watts}}]{ulrich}
\bibinfo{author}{\bibfnamefont{A.~S.} \bibnamefont{Ulrich}},
  \bibinfo{author}{\bibfnamefont{M.}~\bibnamefont{Samia}}, \bibnamefont{and}
  \bibinfo{author}{\bibfnamefont{A.}~\bibnamefont{Watts}},
  \bibinfo{journal}{Biochimica et Biophysica Acta -- Biomembranes}
  \textbf{\bibinfo{volume}{1191}}, \bibinfo{pages}{225} (\bibinfo{year}{1994}).

\end{thebibliography}

\end{document}